**Microtorus: a High Finesse Microcavity with Whispering-Gallery Modes**

Vladimir S. Ilchenko, Michael L. Gorodetsky[†], X. Steve Yao,

and Lute Maleki

Jet Propulsion Laboratory, California Institute of Technology,

MS 298-100, 4800 Oak Grove Dr., Pasadena, CA 91109-8099

[†] Physics Department, Moscow State University, Moscow 119899 Russia

ABSTRACT

We have demonstrated a 165 micron oblate spheroidal microcavity with free spectral range 383.7 GHz (3.06nm), resonance bandwidth 25 MHz ($Q \approx 10^7$) at 1550nm, and finesse $F \geq 10^4$. The highly oblate spheroidal dielectric microcavity combines very high Q-factor, typical of microspheres, with vastly reduced number of excited whispering-gallery (WG) modes (by two orders of magnitude). The very large free spectral range in the novel microcavity - few hundred instead of few Gigahertz in typical microspheres – is desirable for applications in spectral analysis, narrow-linewidth optical and RF oscillators, and cavity QED.

OCIS terms: 060.2340 Fiber optics components; 140.4780 Optical resonators

Microspherical resonators supporting optical whispering gallery (WG) modes have attracted considerable attention in various fields of research and technology. Combination of very high Q-factor ($10^8$-$10^9$ and greater) and sub-millimeter dimensions (typical diameters ranging from few tens to several hundred microns), makes these resonators attractive new components for a number of applications, including basic physics research, molecular spectroscopy, narrow-linewidth lasers, optoelectronic oscillators (OEO), and sensors [1-4]. Effective methods of coupling light in and out of WG modes in microspheres are currently being developed, including single mode fiber couplers and integrated waveguides [5,6].

Whispering-gallery modes are essentially closed circular waves trapped by total internal reflection (TIR) inside an axially symmetric dielectric body. The very high Q of microspheres results from 1) ultra-low optical loss in the material (typically, fiber-grade fused silica), 2) fire-polished surface with Angstrom-scale inhomogeneities, 3) high index contrast for steep reduction of radiative and scattering losses with increasing radius, and 4) two-dimensional curvature providing for grazing reflection of all wave vector components. Grazing incidence is essential for minimizing surface scattering that would otherwise limit the Q far below that imposed by attenuation in the material [7]. For example, in the integrated optical micro-ring and micro-disk cavities [8] based on planar waveguide technology (the light in planar devices is effectively bouncing from flat surfaces under *finite* angle), typical Q-factor is only $10^4$ to $10^5$. The substantially higher Q in the spheres, as compared to micro-disks and micro-rings, comes at the price of a relatively dense spectrum of modes. In ideal spheres, the spectrum consists of $TE(TM)_{lmq}$ modes separated by "large" free spectral range (FSR) defined by the circumference of the sphere, and related to consecutive values of index *l*. In silica spheres of diameter 150 to 400 micron, the "large" FSR should be in the range of 437 to 165GHz, or in the wavelength scale, 3.5 to 1.3nm near the center wavelength 1550nm. Each of $TE(TM)_{lmq}$ modes is ($2l+1$)-fold degenerate with respect to the index *m*. Residual nonsphericity lifts this degeneracy and leads to a series of observable $TE(TM)_{lmq}$ modes separated by "small" FSR in the range 6.8 - 2.5GHz, for the same sphere dimensions, center wavelength, and the eccentricity $\varepsilon^2 = 3\times10^{-2}$, typical of current fabrication methods.

A relatively dense spectrum complicates important applications of microsphere resonators, such as spectral analysis and laser stabilization, and necessitates using intermediate filtering.

In this Letter, we demonstrate a microcavity with a novel geometry that retains two-dimensional curvature confinement, low scattering loss, and very high Q typical of microspheres, and yet approaches spectral properties of the single mode waveguide ring: a highly-oblate spheroidal microcavity, or microtorus.

**Calculation of the spectrum of the dielectric spheroid** (i.e. ellipsoid of revolution) is not a trivial task, even numerically. In contrast to the case of small eccentricity (see, for example, Ref.[9]), analysis in highly-eccentric spheroids cannot be based on simple approximations. Unlike the case of cylindrical or spherical coordinates, orthogonal spheroidal vector harmonics that satisfy boundary conditions may not be derived via a single scalar wave function [10]. Furthermore, the calculation of spheroidal angular and radial functions is also a nontrivial task.

To analyze the properties of eigenfrequencies, we shall rely on physical considerations; however the same approximations can be derived from the asymptotic forms of the spheroidal functions. Very good approximation of WG mode eigenfrequencies in dielectric sphere with radius $a$ much larger than the wavelength can be found as follows:

$$nk_{lmq}a \approx t_{lq} - \frac{\chi}{\sqrt{n^2-1}}, \tag{1}$$

where $t_{lq}$ is the $q$-th zero of the spherical Bessel function of the order $l$ and $\chi = n$ for *TE*-mode and $\chi = 1/n$ for *TM*-mode. For large $l$, $t_{lq} \sim l + O(l^{1/3})$, and it may be calculated either directly or approximated via the zeroes of the Airy function ([11]). The meaning of small second term on the right hand side of Eq.(1) can be understood if we remember that 1) a WG mode is quasiclassically a closed circular beam supported by TIR, 2) optical field tunnels outside at the depth $1/k\sqrt{n^2-1}$, and 3) the tangential component of $E$ (*TE*-mode), or normal of $D$ (*TM*-mode) is continuous at the boundary. Eigenfrequencies of high-order WG modes ($l >> 1; l \approx m$) in dielectric spheres can be approximated via solutions of the scalar wave equation with zero boundary conditions, because most of the energy is concentrated in one component of the field ($E_\theta$ for *TE*-mode and $E_r$ for *TM*-mode).

Based on above considerations, let us estimate WG mode eigenfrequencies in oblate spheroids of large semiaxis $a$, small semiaxis $b$, and eccentricity $\varepsilon = \sqrt{1 - b^2/a^2}$. Since WG modes are localized in the "equatorial" plane, we shall approximate the radial distribution by *cylindrical* Bessel function $J_m(n\tilde{k}_{mq}r)$ with $n\tilde{k}_{mq}a = na\sqrt{k_{lmq}^2 - k_\perp^2} \approx T_{mq}$, where $J_m(T_{mq}) = 0$ and $k_\perp$ is the wavenumber for quasiclassical solution for angular spheroidal functions. For our purposes a rough approximation is enough:

$$k_\perp^2 \approx \frac{2(l-m)+1}{a^2\sqrt{1-\varepsilon^2}} m ;$$

more rigorous considerations can follow the approach given in [12]. Taking into account that $T_{mq} \approx t_{lq} - (l - m + 1/2)$, we finally obtain the following approximation:

$$nk_{lmq}a - \frac{\chi}{\sqrt{n^2-1}} \approx na\sqrt{\tilde{k}_{mq}^2 + k_\perp^2} \approx T_{mq} + \frac{k_\perp^2 a^2}{2T_{mq}} \approx t_{lq} + \frac{2(l-m)+1}{2}\left(\frac{1}{\sqrt{1-\varepsilon^2}} - 1\right) \quad (2)$$

For very small $\varepsilon$, Eq.(2) yields the same value for "small" FSR (frequency splitting between modes with successive $m \approx l$) as earlier obtained by the more rigorous perturbation methods in [9]. In addition, we have compared the prediction of the approximation (2), $l = 100$, with numerically calculated zeroes of the radial spheroidal functions. Even with $\varepsilon = 0.8$, the error is no more than 5% in the estimate of $m$, $m+1$ mode splitting and 0.1% in the absolute mode frequencies. For larger $l$ and smaller $\varepsilon$, the error will evidently be smaller. As follows from (2), with increasing eccentricity, "small" FSR – frequency interval between modes with successive $m$ – becomes compatible with the "large" FSR: $(k_{l+1mq} - k_{lmq}) \sim (k_{lm+1q} - k_{lmq}) \sim k_{lmq}/l$. In addition, excitation conditions for modes with different $m$ become more selective: e.g. optimal angles for prism coupling vary with $\varepsilon$ as $\frac{k_\perp}{k_{lmq}} \propto (1-\varepsilon^2)^{-1/4}$.

In our **experiment**, we prepared a spheroidal (microtorus) cavity by compressing a small sphere of low-melting silica glass between cleaved fiber tips. The combined action of surface tension and axial compression resulted in the desired geometry (see typical microcavity in Fig.1). One of the fiber "stems" was then cut and the whole structure was installed next to a standard prism coupler. The WG mode spectrum was observed using a DFB laser near the wavelength of 1550nm. The laser was continuously frequency-scannable within the range of ~80GHz (by modulating the current), and temperature tunable between 1545.1 to 1552.4 nm. A high-resolution spectrum over 900GHz was compiled from 15 individual scans with 60GHz increments obtained by changing the temperature. The frequency reference for "stitching" the spectral fragments was obtained by recording the fringes of high-finesse ($F \sim 120$) Fabry-Perot ethalon (FSR=30GHz) and additional frequency marks provided by 3.75GHz amplitude modulation (Fig.2). Total drift of the FP was less than 400MHz over the full 15-minute measurement time. The compiled spectrum is presented in Fig.3. The spectrum is reduced to only two whispering-gallery modes of selected polarization excited within the "large" free spectral range of the cavity FSR = 383GHz, or 3.06nm in the wavelength domain. The transmission of "parasitic" modes is at least 6dB smaller than that of the principal ones. With individual mode bandwidth of 23MHz, the finesse $F = 1.7 \times 10^4$ is therefore demonstrated with this micro-resonator. We can compare this result with the predictions of our approximate expression (2). For dimensions of our cavity $a = 82.5\mu m$, $b = 42.5$ μm ($\varepsilon = 0.86$) and the refraction index of the material n = 1.453, the principal mode number for *TE*-modes at the wavelength 1550nm should be $l \approx 473$ and the "large" FSR

$$\nu_{lmq} - \nu_{l-1,mq} = \frac{c}{2\pi na}(t_{lq} - t_{l-1,q}) = \frac{c}{2\pi na}(1 + 0.617 l^{-2/3} + O(l^{-5/3})) \approx 402 GHz .$$

The "small" free spectral range should be

$$\nu_{lmq} - \nu_{l,m-1,q} = \frac{c}{2\pi na}(\frac{1}{\sqrt{1-\varepsilon^2}} - 1) \approx 382 GHz$$

In the experimental spectrum, frequency separation between the largest peaks 1 and 2 in Fig.3 is equal to 383.7 ± 0.5GHz. It may therefore be attributed as corresponding to the small FSR, in good agreement with the above estimate, if we take into account an approximately 2% uncertainty in the measurement (limited mainly by the precision of geometrical evaluation of cavity dimensions). The separation between the peaks 3 and 4 in Fig.3, is equal to 400.3 ± 0.5GHz, and is in turn close to the estimate of "large" free spectral range.

It may be argued that despite the large splitting between the modes with adjacent values of index *m* (of the order of "large" FSR), one may still expect a dense spectrum resulting from the overlap of many mode families with different principal number *l*. In practice, however, it is exactly the coincidence in the frequency domain of WG modes with different main index *l,* and rapidly increasing difference *l – m,* that should be responsible for effective dephasing of the "idle" modes from the evanescent coupler, resulting in the reduction of modes in the observed spectrum. A complete analytic interpretation is beyond the format of this Letter and should include a wider range spectral mapping, a study of microcavities of different eccentricity, the calculations of field distribution, and analysis of phase-matched excitation in the evanescent coupler.

In **conclusion**, experimental results indicate a substantial reduction (up to 2 orders of magnitude) in the number of excited WG modes in highly oblate spheroids compared to typical microspheres. This reduction in the mode density is obtained without sacrificing the high Q associated with these structures. The novel type of optical microcavity demonstrates a true finesse on the order of $10^4$, a free spectral range of the order of few nanometers, and a quality-factor $Q \sim 1\times10^7$. Based on these results, we conclude that a complete elimination of "transverse" WG modes maybe expected in spheroidal cavities of higher eccentricity. Further increase of the Q and the finesse may also be expected with a refinement of the fabrication technique. The decrease in the density of mode spectrum of ultra-high-Q microcavities offers new applications in laser stabilization, microlasers, various passive photonics devices, and in fundamental physics experiments.

FIGURE CAPTIONS

Fig.1. Photograph and the schematic of the microcavity geometry. Near the symmetry plane (at the location of WG modes), toroidal surface of outer diameter $D$ and cross-section diameter $d$ coincides with that of the osculating oblate spheroid with large semiaxis $a = D/2$ and small semiaxis $b = \frac{1}{2}\sqrt{Dd}$

Fig.2. Schematic of the experimental setup to obtain wide range (~900GHz, or 7.2nm) high-resolution spectra of WG modes in microcavity

Fig.3. Spectrum of *TE* whispering-gallery modes in spheroidal dielectric microcavity ($D = 2a = 165\mu m$; $d = 42\mu m$; $2b = 83\mu m$). Free spectral range (frequency separation between largest peaks 1 and 2) 383.7GHz (3.06nm) near central wavelength 1550nm. Individual resonance bandwidth 23MHz (loaded Q = $8.5\times10^6$). Finesse $F = 1.7\times10^4$

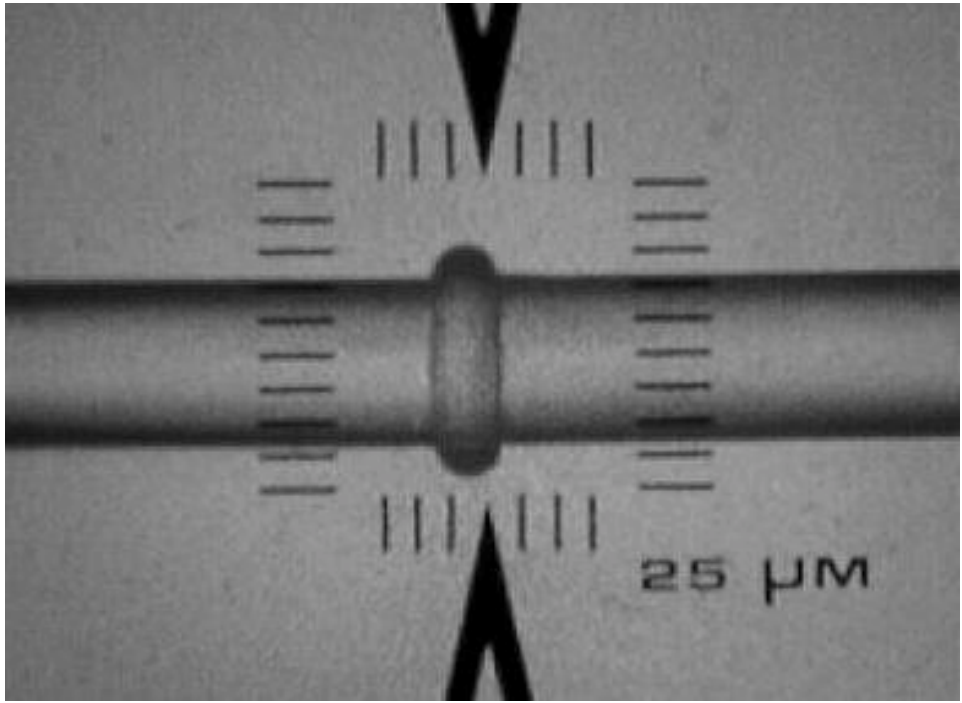

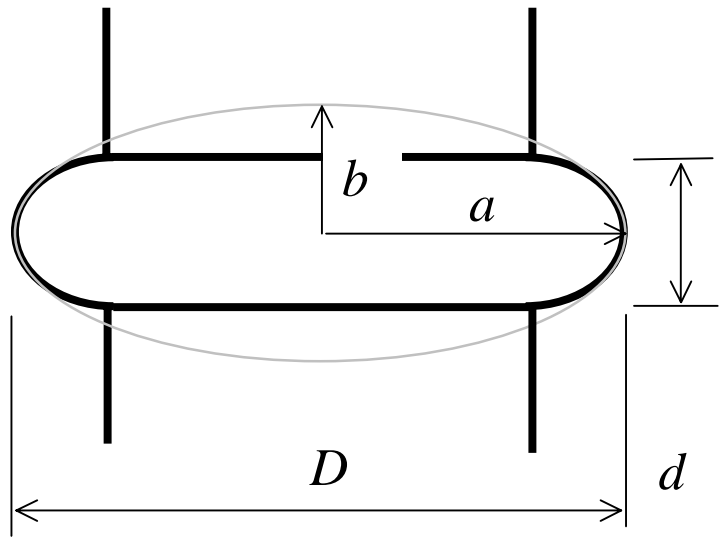

Fig.1. V.S.Ilchenko et al., "Microtorus: a High Finesse…"

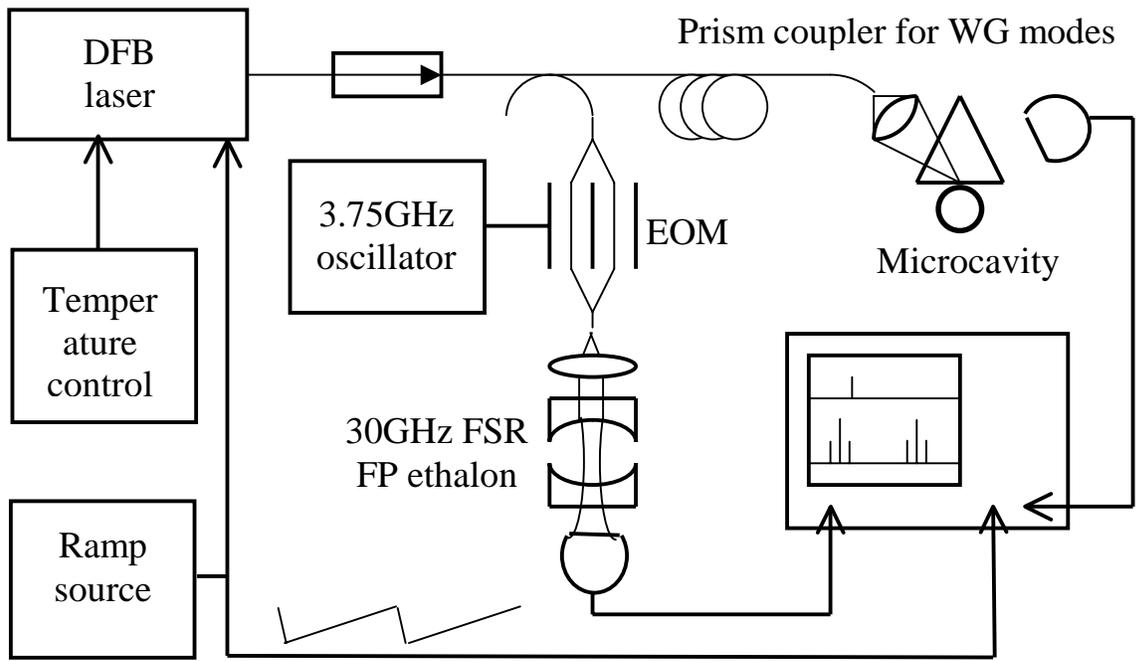

Fig.2. V.S.Ilchenko et al., "Microtorus: a High Finesse…"

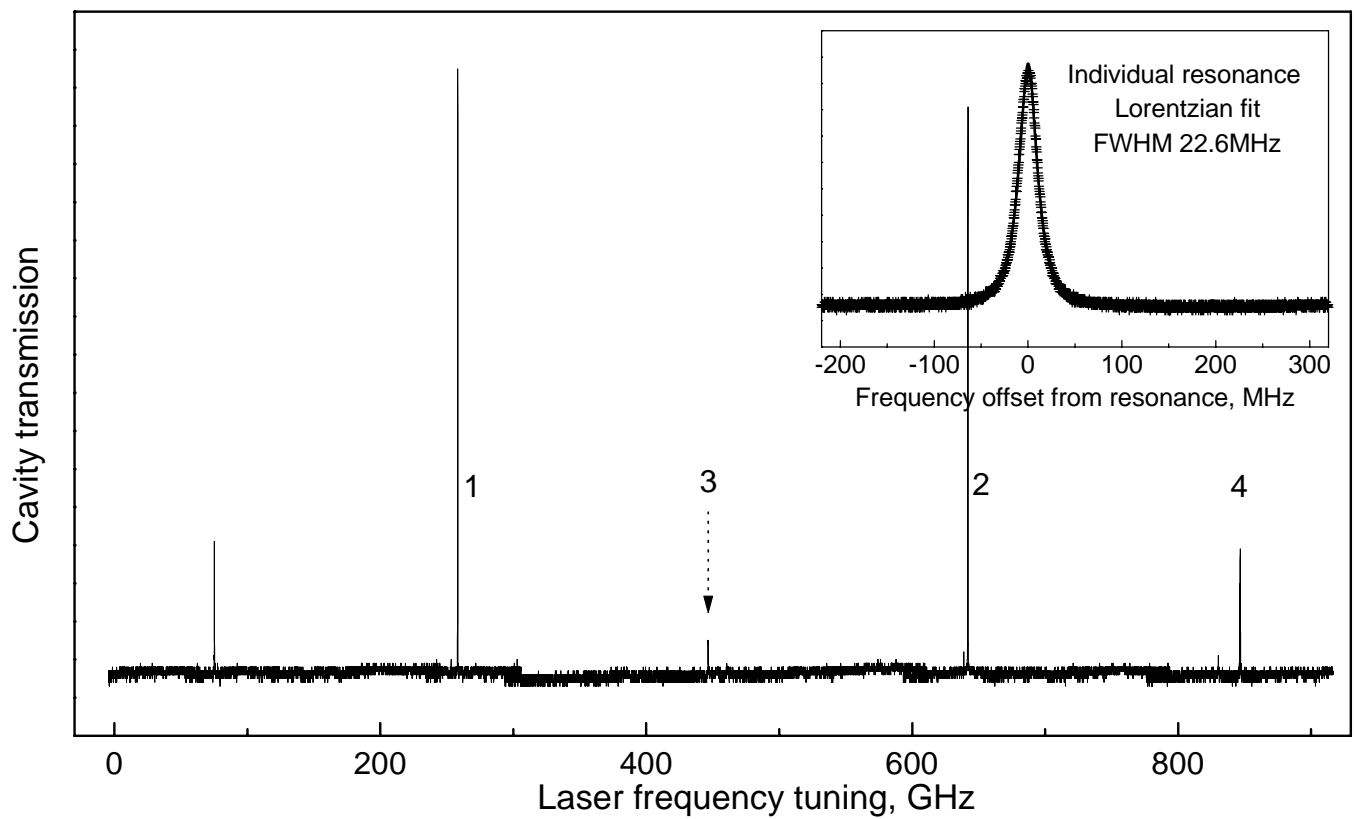

Fig.3. V.S.Ilchenko et al., "Microtorus: a High Finesse…"